# MICROBEAM AND PULSED LASER BEAM TECHNIQUES FOR THE MICRO-FABRICATION OF DIAMOND SURFACE AND BULK STRUCTURES

S. Sciortino[1,2], M. Bellini[3,4], F. Bosia[5,6], S. Calusi[1,2], C. Corsi[3], C. Czelusniak[1,2], N. Gelli[2], L. Giuntini[1,2], F. Gorelli[3,4], S. Lagomarsino[1,2], P. A. Mandò[1,2], P. Olivero[5,6,7], G. Parrini[1], M. Santoro[3,4] A. Sordini[8], A. Sytchkova[9], F. Taccetti[2], M. Vannoni[8]

[1] *Dipartimento di Fisica e Astronomia, Università di Firenze, Via Sansone 1, I-50019, Sesto Fiorentino, Firenze, Italy*
[2] *Istituto Nazionale di Fisica Nucleare (INFN), Sezione di Firenze, Via Sansone 1, I-50019, Sesto Fiorentino, Firenze, Italy*
[3] *European Laboratory for Non-Linear Spectroscopy, Via Nello Carrara 1, 50019 Sesto Fiorentino (FI), Italy*
[4] *Istituto Nazionale di Ottica (INO-CNR), Largo Enrico Fermi 6, 50125 Firenze (FI), Italy*
3INFN, Sezione di Pisa, Largo B. Pontecorvo 3, I-56127, Pisa, Italy
[5] *Physics Department and "Nanostructured Interfaces and Surfaces" inter-departmental centre, University of Torino, via P. Giuria 1, 10125 Torino, Italy*
[6] *INFN Sezione di Torino, via P. Giuria 1, 10125 Torino, Italy*
[7] *Consorzio Nazionale Interuniversitario per le Scienze Fisiche della Materia (CNISM), via della Vasca Navale 84, 00146 Roma, Italy*
[8] *CNR, Istituto Nazionale di Ottica (INO), Largo E. Fermi 6, 50125 Arcetri, Firenze, Italy*
[9] *ENEA Optical Coatings Group, via Anguillarese 301, 00123 Rome, Italy*

*Abstract*

Micro-fabrication in diamond is involved in a wide set of emerging technologies, exploiting the exceptional characteristics of diamond for application in bio-physics, photonics, radiation detection. Micro ion-beam irradiation and pulsed laser irradiation are complementary techniques, which permit the implementation of complex geometries, by modification and functionalization of surface and/or bulk material, modifying the optical, electrical and mechanical characteristics of the material.
In this article we summarize the work done in Florence (Italy) concerning ion beam and pulsed laser beam micro-fabrication in diamond.

**Introduction**

Microbeam ion implantation in the MeV range and laser micro-fabrication are techniques exhibiting largely complementary features. The relatively short range of MeV implantation makes it useful for the fabrication of structures parallel to the surface of the sample at depths from a few to tens of micrometers, with a vertical resolution limited by the width of the Bragg peak and a lateral one better than one micrometers in the most recent high performance setups [1]. The types of structural modification allowable by ion implantation range from electrical [2] to optical [3-5], mechanical and chemical characteristics [6]. Laser material engineering, on the other hand, depending on wavelength, energy and pulse width, is useful in ablation or amorphization of the material [7,8], and is suitable for the modification of the surface or of interior of the sample (up to centimeters, theoretically), with a lateral resolution comparable to that of the microbeams but with a vertical definition (in the bulk) limited by the focusing aperture to about ten micrometers.
In diamond, these two techniques could pave the way to the integration of micro-devices with applications in particle detection, bio-sensing, micro-optics and quantum-optics. Both ion damaging (followed by appropriate annealing [9,10]) and sub-bandgap pulsed laser irradiation are capable of increase the conductivity of the material by modification of the bonding hybridization, from $sp^3$ to $sp^2$ [11]. Thus, micro-beam writing can be employed in the fabrication of conductive channels or pads under the surface of diamond, while pulsed laser graphitization is suitable for fabrication of conductive columns perpendicular to the surface or of conductive channels at the surface level. In this way, electrodes inside diamond can be implemented in three-dimensional diamond detectors, or in micro-electrodes arrays employed in studies on biological tissues, or in Stark-effect tuned optical micro-cavities, just to mention some of the possible applications. Moreover, the optical modification of the material induced by ion implantation can be used to implement light guides in micro-optical devices. Doping by ion implanting can be employed both in tailoring the band-gap of diamond and in deterministic implantation of color centers for quantum applications. On the other hand, laser ablation and microbeam graphitization, followed by chemical etching, are useful to model the surface of the material for applications ranging from bio-physics to optics.

For all these applications, ion beams of different species and at different current levels are needed together with different types of pulsed laser beams. At the LABEC laboratories of Florence, Italy, we can employ two lines of a 3 MV tandem accelerator: the external microbeam setup with a lateral resolution of 10-20 μm for modification of the optical and electrical properties of the material [12], and the pulsed laser beam facility for very low-current level implantations [13]. At the LENS laboratories, also in Florence, a pulsed laser apparatus is arranged with two different laser lines on a same optical setup: a 30 fs, 800 nm Ti:Sapphire laser and an 8 ns, 1064 nm Nd:YAG laser source, both operating in the microjoules-per-pulse range [11]. Several techniques are employed to characterize the artifacts: electrical characterization at the laboratories of INFN (Florence), geometrical profiles and refractive index at the at the INO laboratories (Florence), Raman characterization at the LENS laboratories.

In this article, we review the work done in Florence in the micro-modification of the structural properties of diamond (INFN, Department of Physics of Turin, CNR of Rome collaboration). Work has been done in microbeam modification of the optical properties of diamond [14-18], microbeam writing of optical waveguides in the bulk diamond [16,19], pulsed laser fabrication of buried and superficial conductive channels [11,20], fabrication of three-dimensional diamond particles detectors [21]. All the expertise acquired in the fabrication and characterization of micro-structures in diamond can be considered ready to use for the realization of diamond integrated devices.

**1. Modification of the complex refractive index due to ion implantation**
In this section we report on the refractive index modification of high quality, chemical vapour deposited IIa diamond samples, irradiated with 2 and 3 MeV protons.

**Ion implantation**
The diamond samples were implanted at the external scanning microbeam facility [25] of the 3 MV Tandetron accelerator of the INFN LABEC Laboratory in Florence. The sample to be implanted was kept out of vacuum, thus allowing its easy handling, positioning and monitoring [26].

Proton beams were focused on the polished side of the samples to a spot of around 10 μm (3 MeV) or 20 μm (2 MeV) FWHM. Different zones of the samples were implanted at fluences ranging from $10^{15}$ /cm$^2$ to $10^{17}$ /cm$^2$ .

The overall precision on the implanted charge determination is about 1%. Possible systematic errors in the charge determination, affecting all the experimental points with a common scale factor, amount to 10% of the measured value. After ion implantation, the size of the irradiated area was measured on the *OPD* maps as described below, the resulting precision on the area determination is about 2%.

**Measurement of the *OPD* and *ALD***

In order to evaluate the *OPD* due to ion-induced damage, the phase shift of a laser beam crossing the damaged diamond layer was determined using a commercial laser interferometric microscope (Maxim 3D, Zygo Corporation, Middlefield, CT, USA) with a 20 × micro-Fizeau objective, operating in the $\lambda_{He-Ne}$ = 632.8 He-Ne laser line, with horizontal and vertical resolutions of 1.68 μm and 0.63 nm, respectively, and with a field view of 349 × 317 μm [14].

A He-Ne laser beam is properly expanded to invest the full area of the sample; the micro- Fizeau objective contains a beam-splitter that reflects part of the light ("reference beam"), while the remaining part crosses the sample and is reflected from a high-quality external mirror ("test beam"). The diamond is slightly tilted to avoid undesired internal reflections between the two opposite surfaces of the sample. The interference pattern of the reference and test beam is recorded by a CCD camera.

Using the phase shift method [28] it is possible to reconstruct the relative phase Δ of the test beam at each pixel: the contributions of the beam splitter and the high-quality mirror is accounted for and removed. The phase difference Δ reflects the optical path difference: $\Delta = \dfrac{2\pi}{\lambda_{He-Ne}}$

The absorption length difference was evaluated, for each implantation, by the ratio between the transmittance $T_0$ of the unimplanted substrate and the value $T$ measured through a chosen damaged area:

$$ALD = \dfrac{\lambda}{4\pi} \log\left(\dfrac{T_0}{T}\right)$$

The transmittance spectra were acquired with a setup described in Ref. [15].

Both the *OPD* and the *ALD* measurements are affected by swelling, i.e., the expansion of the implanted material, which determines both a further phase shift of the probe laser beam and an additional absorption contribution. Nevertheless, since the gradient of the displacement of each layer in diamond $\dfrac{dz'}{dz}$ and the relative variation of the refractive index $\dfrac{\Delta n}{n}$ are both small with respect to unity, it can be shown [16] that the values of *OPD* and *ALD* due to the variation of the refractive index alone can be obtained by the measured ones ($OPD_m, ALD_m$), by the simple equations:

$$OPD = OPD_m - (n_0 - 1)h$$
$$ALD = ALD_m - \kappa_0 h$$

where $h$ is the swelling height. This parameter has been measured by means of a white-light interferometry microscope (Newview, Zygo Corporation).

In our measurements, the product $k_0 h$ is negligibly small (well below 0.1%) and its contribution has been neglected, but the product $(n_0 - 1)h$ amounts to about 15% of the measured OPD, and it has been properly subtracted.

**Results and Discussion**

A linear model has been exploited [17] to interpret the OPD and ALD measurements in terms of the modification of the real and of the immaginary part of the refractive index, taking into account the damage profile produce by 2 and 3 MeV protons and calculated by means of a Monte Carlo SRIM simulation. It results that both the OPD and the ALD are linear in the ion fluence and are proportional to the average number $I^E$ of vacancies produced by each ion of specific energy $E$. Figure 1 shows how the ratio OPD/$I^E$ is proportional to the fluence and independent on the energy; for the ALD, a similar plot has been obtained [17].

It results that the ion induced complex refrative index, for fluences up to the highest reached in our experiments, can be expressed as:

$$\bar{n} = 2.41 + [(4.84 \pm 0.05) + i(2.86 \pm 0.04)] \times 10^{-23} \text{ cm}^3 \rho \quad (1)$$

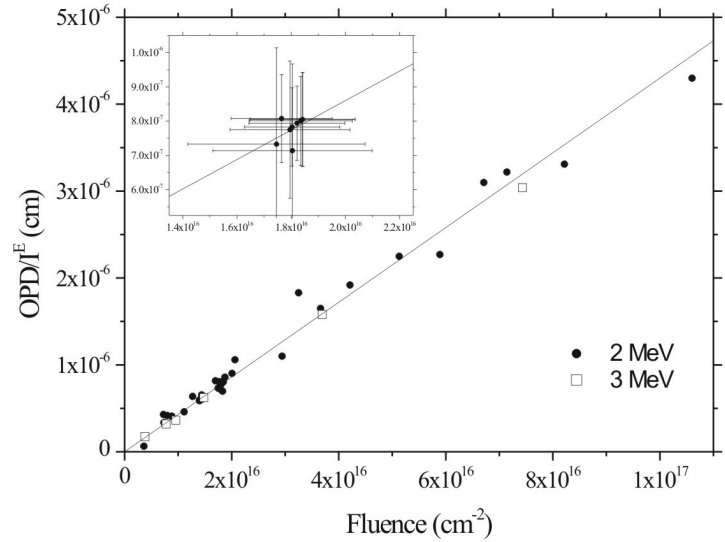

Figure 1. Linear trend of the *OPD*( *E*, ϕ)/ *I*( *E*) ratio as a function of the fluence ϕ. In the inset, particular of the points representing eight different implantations at a same nominal fluence but with different values of the instantaneous current (a factor 5 of variation).

Where ρ is the vacancy density produce by the irradiation in vac/cm$^3$. The experimental results point out that the variation of the refractive index depends only on the overall vacancy density induced by the radiation during the process, irrespectively of the ion energy and of the beam intensity. Previous reports about the optical characteristics of ion-damaged diamond [26-29] also report increasing trends of the real part of the refractive index. The linear coefficients, although determined with much higher uncertainty, are compatible with the results summarized by eq. 1. In a very early report [29], the refractive index of diamond implanted with 20 keV C$^+$ ions exhibits a monotonic increase as a function of implantation fluence, with linear coefficients strongly dependent on the measured sample and ranging from about 2 to 10×10$^{-23}$ cm$^3$.

The linear dependence holds up to a damage level at which the refractive index seems to saturate; such saturation level corresponds to a total atomic concentration of $4.5\times10^{21}$ vacancies cm$^{-3}$, a value slightly exceeding the maximum damage density explored in the present work ($2.5\times10^{21}$ vacancies cm$^{-3}$). For one of the four diamond samples reported in [29] (sample I), the dependence of the refractive index from the damage density is in very satisfactory agreement with our result, while other samples exhibit rather different trends. From such a very early report it is not possible to reconstruct the types of the different diamond samples employed.

Differently from what reported in [33], in [31] no clear trend emerges in the variation of the refractive index and therefore a direct comparison with the present work is difficult. In [32] the authors report a low value of the refractive index for the heavily damaged buried layers, whose damage-induced vacancy density amount to about $4\times10^{22}$cm$^{-3}$. In these conditions, the degree of amorphization/graphitization exceeds by far what reported in the present work. Finally, it is worth remarking that the results of this work are in good agreement with recent ellipsometric studies of the refractive index variation in shallow layers implanted with 180 keV B ions, for which consistent linearly increasing trends are reported in the at low damage density regime [30]. In particular, at wavelength 632.8 nm , a linear coefficient of $(3.8 \pm 0.3)\times10^{-23}$ cm$^{-3}$ can be obtained for the dependence of the real part of the refractive index, in satisfactory agreement with the value reported in this work, particularly if it is considered that different implantation conditions and analytical techniques were employed.

The increasing trend of the refractive index as a function of induced damage is somewhat surprising with respect to what reported in other materials, such as quartz [33] or zircon [34], for example. This is because the most direct effect of ion implantation in crystals usually consists in the progressive amorphization of the substrate, which invariably leads to a decrease of the atomic density and therefore of the refractive index. Although often quantitatively predominant, the above-mentioned process is not the only effect determining a variation in refractive index. Beside volume expansion, other damage-related effects can occur which have a significant and direct effect on the refractive index, namely changes in atomic bond polarizability and structure factors, as expressed by the Wei adaptation of the Lorentz-Lorenz equation [35]:

$$\frac{\Delta n}{n} = \frac{(n^2-1)(n^2-2)}{6n^2}\left(-\frac{\Delta V}{V}+\frac{\Delta \alpha}{\alpha}+F\right)$$

where $V$ is the volume, $\alpha$ is the polarizability and $F$ is the structure factor of the target implanted material.

Although the volume expansion term is dominating in most cases, the structural modification results in changes of the chemical bonds and subsequently of the material polarizability. Such changes can be either positive or negative in sign and, therefore, it is reasonable to expect strong polarizability-related effects in a peculiar material such as diamond, in which the nature of the chemical bond can be subjected to drastic changes (i.e. from the strongly covalent $sp^3$ bonds to $sp^2$ bonds).

While for low damage levels (well below the amorphization threshold, as mentioned above), polarizability-related effects related to the formation of isolated $sp^2$ defects can dominate over the volume effects, it is reasonable to expect that at higher damage levels the amorphization of the diamond $sp^3$ lattice can lead to predominant density effects and thus to the reduction of the refractive index, as indeed observed in [32].

We conclude by remarking that further investigation should be necessary to ascertain if the same mechanisms occur also for the damage induced by other ion species, but the present work indicates that a proton beam can be used in tailoring the optical properties of diamond in the MeV range with the help of a common damage simulation software such as SRIM. The methodology of measurements and analysis which adopted for this study is of ease and versatile use, for application for any transparent material within very large range of energies and fluences.

## 2. Waveguides engineering in single crystal diamond by MeV proton implantation

**Ion implantation of the waveguides**

To perform this study, three surfaces of a IIa monocrystalline CVD diamond were optically polished to a roughness of 1 nm: the two opposite $3.0 \times 3.0$ mm$^2$ faces and one of the four lateral $3.0 \times 0.5$ mm$^2$ faces down to a roughness of some nanometers. To obtained controlled increments of the refractive index a 3 MeV proton beam was focused on the small polished side of the sample to an approximately Gaussian spot 12 µm wide, and scanned along a 500 µm rectilinear path perpendicular to the large polished faces (longitudinal direction of the guide, see the schematics of Fig. 2 [19]). The fluences were $2 \times 10^{16}$, $1 \times 10^{16}$, $5 \times 10^{15}$ cm$^{-2}$ in the central region of each implantation, with an estimated uncertainty not exceeding 5%. The resulting vacancy density distribution, as calculated using SRIM Monte Carlo simulations, follows the characteristic distribution, also recalled in Fig. 2 (left panel), peaked at a depth of approximately 50 µm.

**Optical characterization and interpretation of data**

The as-prepared structures were then observed with the Maxim inteferometer, previosly used to characterize refractive index variations by measuring the *OPD*. In this case the phase maps obtained with the micro-inteferometer can be interpreted as a direct measurement of the amplitudes of the modes propagating along the guide.

In fact, as the structures under consideration have a cross-sectional dimension comparable to that of the wavelength of the radiation, the radiation emerging from the diamond will be given by a principal plane-wave part plus a perturbation produced by the structures themselves. Consequently, the field will be given by the sum of contribution from which the amplitude map of the mode can be obtained as a sum of different simultaneous modes propagating in the waveguide.

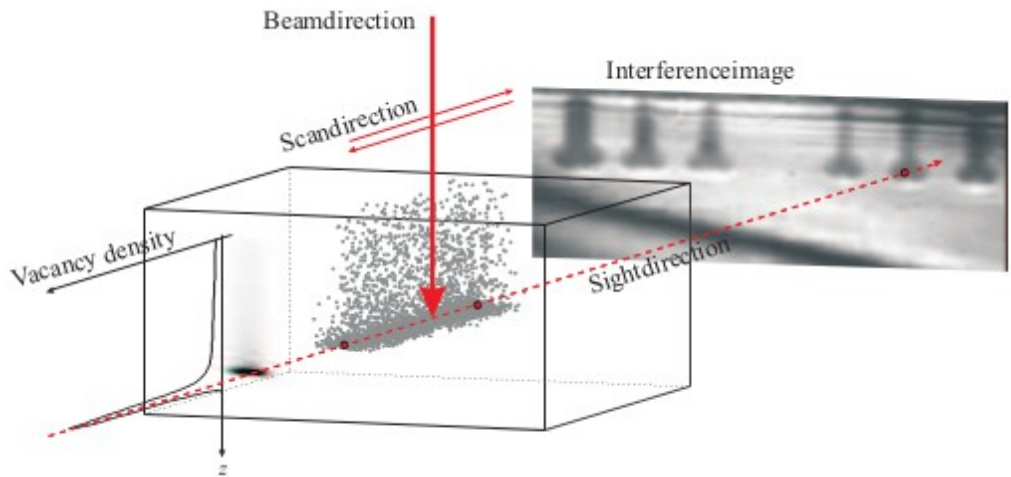

Figure 2. Schematics of the implantation geometry and the resulting interference pattern. Implantation fluences from left to right: $2 \times 10^{16}$ cm$^{-2}$ (one implantation), $1 \times 10^{16}$ cm$^{-2}$ (two implantations), $5 \times 10^{15}$ cm$^{-2}$ (the last three implantations).

For the calculation of the field modes, a 2-dimensional finite element model (FEM) of the irradiated regions was employed, taking into account the local modifications in the refractive index induced by proton damage, quantified in terms of the induced vacancy density and calculated by means of a Monte Carlo simulation (SRIM). Once given the vacancy density at every cell of the simulation grid, the local variation of refractive index at the He-Ne wavelength of 632.8 nm is calculated from the simple relation (1).

Then, the experimentally obtained phase maps were compared with a superposition of the calculated amplitude maps, by fitting them with a linear combination of the propagating modes. Since the relative amplitudes of the modes excited in the waveguides depend in a sensitive way from the illumination conditions, different positions of the sample on the focal plane may imply different weights to be assigned at each particular mode. In Figure 3 different images of the implantations at fluencies
of $2 \times 10^{16}$, $1 \times 10^{16}$ and $0.5 \times 10^{16}$ cm$^{-2}$ are shown along with the best fit obtained with 30 different propagation modes (ten for each structure) and two plane sinusoids, taking into account the reflections on the two planes. It is evident that the same set of propagation modes, although with different weights, fits the different images. From the inspection of these images we conclude that the adherence of the fit to the experimental two-dimensional profiles is very good in the cap layer between 0 and about 45 μm in depth, where the relative damage is small, while at end-of-range the structures seems to be more diffuse, probably due to the distortion induced by diffraction on the highly opaque regions, in correspondence with the considered structures.

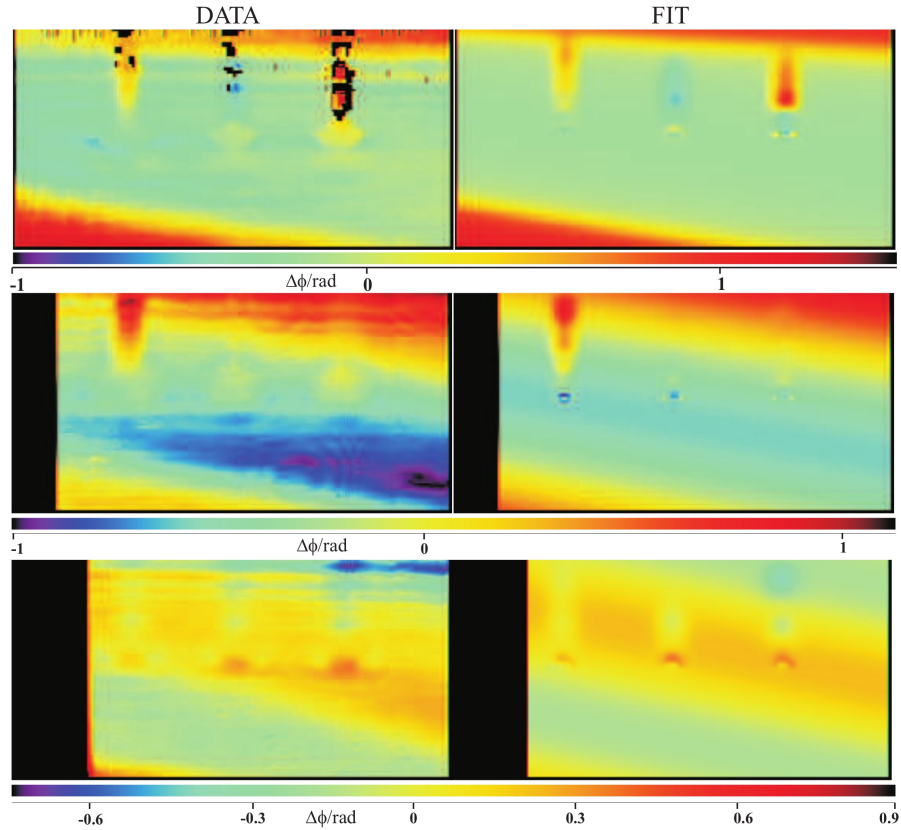

Figure 3. Comparison of the measured phase shift maps (left) and of the fit (right) obtained by linear superposition of modes amplitudes and a background taking into account multiple reflections effects. Top and middle: images obtained from three adjacent guides irradiated at $2 \times 10^{16}$ cm$^{-2}$ (the left one) and at $1 \times 10^{16}$ cm$^{-2}$ (the others). Bottom: images obtained by three equally irradiated guides at a fluence of $5 \times 10^{15}$ cm$^{-2}$.

## 3 Laser graphitization of diamond

In this section we describe surface and bulk laser graphitization of diamond aimed to fabricate (three-dimensional) diamond-based radiation detectors. The experimental setup described in [11] employs two pulsed laser sources:
a) a Nd:YAG Q-switched source with an 8 ns pulse width, 1064 nm wavelength, pulse energies in the range 10–60 μJ and repetition rates from 1 to 10 kHz.
b) a Ti:sapphire femtosecond laser source of 30 fs pulse width, 800 nm wavelength, pulse energies between 3 and 18 μJ and repetition rate of 1 kHz.
Both beams have been focused either on the diamond surface or in the diamond bulk. The samples used were Element Six high-purity monocrystalline $4.5 \times 4.5 \times 0.5$ mm$^3$ and polycrystalline $5 \times 5 \times 0.5$ mm$^3$ CVD diamond plates.
The graphitic structures we implemented are:
A) superficial conductive tracks obtained by keeping the front surface of the diamond in the focal plane of the objective and translating it at constant velocity (xy- directions).
B) Buried conductive wires obtained by focusing the laser beam on the back diamond surface and moving the focus at constant velocity perpendicularly to the surface, across the bulk for 100– 500 μm (z-direction).

## Structural and electrical characterization

Only the ns-pulsed laser source appears to be useful in fabrication of superficial conductive tracks, because the fs-laser source causes ablation of diamond, and leaves only a very thin layer of modified material. On the contrary, the ns-laser source creates deep (up to 50 µm) and narrow (~ 10 µm) channels uniformly filled with an opaque material, which results ablated only for a depth from 3 to 7 µm.

The depth of the channels increases with the number of laser pulses (up to about 50 µm at about 700 pulses/ point) , on the contary it is quite independent on the pulse energy (at least up to 50 µJ/pulse), provided that the energy lies above a threshold of about 6 µJ/pulse. This is the threshold found if the irradiation starts from a zone where the material is already graphitized, while if the graphitization has to start from undamaged diamond the threshold is placed at about 37 µJ/pulse.

The resistivity of the modified material, as measured on different tracks, fabricated with different energy-per-pulse and number of pulses-per-point, is 8 ± 4 mΩcm, which is not so far from those reported for amorphous graphite with no clear dependence on the process parameters.

Raman characterization confirms that the modified material consists in a phase of disordered $sp^2$ carbon [11]. we found invariantly a feature with two wide peaks: one centered at 1580 cm$^{-1}$ (G peak of graphite) and one whose position depended on the excitation wavelength, identified as the D peak of disordered graphite [36].

Both the sources we employed are capable to write buried conductive channels perpendicular to the beam entrance surface of diamond, but with different geometrical and physical characteristics. The cross-sectional area of both types of structure depends on the pulse energy, being roughly proportional to the difference between the pulse energy and a threshold value which is about 2 µJ for the fs-pulsed laser source and 9 µJ for the ns-one. In the case of ns-pulsed laser, in order to grow a buried column with such a low value of the energy- per-pulse, it is necessary to initiate it on an already graphitized zone on the back side of the diamond sample. The morphological characteristics of the two kinds of columns are quite different: ns-laser made structures are quite irregular in cross- section and exhibit cracks which are more and more evident as the value of the energy-per-pulse increase. On the contrary, fs-laser made columns are more regular in section and show traces of ruptures only for very high values of the energy-per-pulse employed. The two types of wires also exhibit a very different electrical behavior. The mean resistivity obtained for the ns-source wires was about 60 mΩcm, while that for the fs-source wires was an order of magnitude greater (about 900 mΩcm) in agreement with Kononenko et al. [37]. The Raman spectra of the buried structures were observed through the lateral polished surface of the diamond plate, at a distance of about of 40 µm from the graphitic column.

The Raman analysis of the two kinds of structures explains the difference in their electrical behavior. The 1332 cm$^{-1}$ line of diamond is superimposed to the D peak, due to the 40 µm-thick layer of diamond in front of each column, and a distinct G peak at 1580 cm$^{-1}$ is clearly observable. Moreover, a feature at 1090 cm$^{-1}$ is seen, in the structures fabricated with the ns-pulsed laser source, around the graphitic structures within a distance of a few micrometers.This peak is attributed to nano-crystalline diamond [38], or to Z-carbon [39], an $sp^3$ phase which is stable at pressures exceeding about 9.8 GPa. The local pressure has been determined from the stress-induced deformation of the diamond line at 1332 cm$^{-1}$.

A quantitative analysis was carried out taking as an index $r$ of the graphitic content of the graphitic structures the ratio between the G-peak area and the area of the 1332 cm$^{-1}$ peak of unmodified diamond at the same depth. Bidimensional maps of the graphite contents in the modified regions were derived from this analysis[11]. It can be observed that the maximum $r$ index measured in the structures created with the nanosecond laser source is one order of magnitude larger than that of the femtosecond structures. Therefore the resistivity values of differently fabricated structures are related to the different content in graphite of the material. In both cases we interpret these results in terms of a mixture of two phases in which conduction takes place by percolation between graphite micro or nano-crystals dispersed in an $sp^3$ matrix.

Bidimensional maps of the pressure gradient in the modified region of the graphitic channels was derived from the analysis of distortion/shift of the diamond Raman line [40] .

From the maps it becomes apparent that the regions occupied by the graphitic phase and by the $sp^3$ nanostructured phase are related to a compressive stress in the diamond around them which can be as high as 10 GPa, not so far from the maximum pressure for which graphite is stable at the thermodynamic equilibrium, that is the graphite–diamond–liquid triple point pressure, at about 13 GPa [41]. This explains the reduced graphitic content and the high values of resistivity of the buried material. The very high elastic constants of diamond and graphite and the low density of graphite with respect to diamond would determine, in the case of a complete transformation of diamond in graphite, very high pressure of the buried graphitic phases, which can be estimated in about 60 GPa. But graphite is stable at the thermodynamic equilibrium only below about 13 GPa. Consequently, only a high density mixed phase can crystallize, in a way that the local pressure never exceeds, after the phase formation, those permitted by thermodynamics. A high density phase can be obtained only in a material relatively poor of $sp^2$ bonds, determining an intrinsic higher resistivity of the buried graphitic electrodes with respect to the surface ones.

**Three-dimensional diamond detectors**

The concept of three-dimensional detectors has been conceived for silicon detectors [22] in order to improve the radiation resistance of solid state detectors. In the last years the concept has been also applied to diamond [21, 23, 24], exploiting the pulsed laser writing techniques made available in the meantime, mainly for optical applications [8].

We fabricated different sensors made on monocrystalline and polycrystalline high purity CVD 0.5 mm thick diamond [21]. The geometry of all the 3D sensors fabricated are based on the repetition of "elementary cells" in which two oppositely polarized columns lie, respectively, at a vertex and at the center of the cell. The dimensions of the elementary cell was from $70\times114$ μm$^2$ to $100\times160$ μm$^2$. The diameter of each column is about 10 μm and 5 μm for the fs-laser-made columns and for the ns-ones, respectively. Reference structures were also fabricated, implementing with the ns laser two graphitic combs with a pitch of 80 μm on the two sides of the samples, without buried columns, in order to compare the performances of the 3D structures with a conventional planar sensor. Fig. 4 shows an image of four different sensors fabricated on a single crystal diamond.

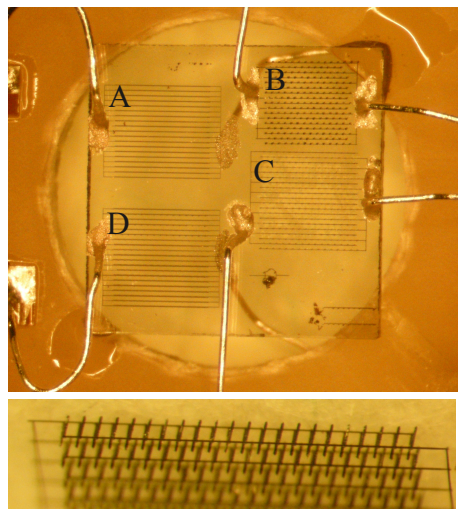

Figure 4. Top. four different sensors fabricated on a single crystal diamond A:Reference planar sensor; B: fs-made sensor; C-D ns-made sensors; D: OSC ns-made sensor. Bottom. Detail of a 3D fs-made sensor.

The collection efficiency of the sensors to relativistic beta particles has been measured using a setup described in detail in ref. [42]. In Figure 5 the dependence of the average signal on the bias voltage is shown for the reference and for the 3D fs-made sensor in the monocrystalline sample. The figure also shows the statistical distribution of the signals for the two sensors at saturation. Full collection (19000 electrons) occurs for both sensors, confirming that superficial graphitic electrodes fabricated with the nanosecond laser source do not exhibit signal loss (see also Ref. [20]) and demonstrating as well that the femtosecond buried columns are suitable electrodes for charge collection. Moreover signal saturation for the 3D sensor (which depends on the applied electric field) occurs at a bias voltage one order of magnitude lower than that of the reference sensor. This confirms that charge transport takes place between electrodes whose interdistance is much lower than the sensor thickness.

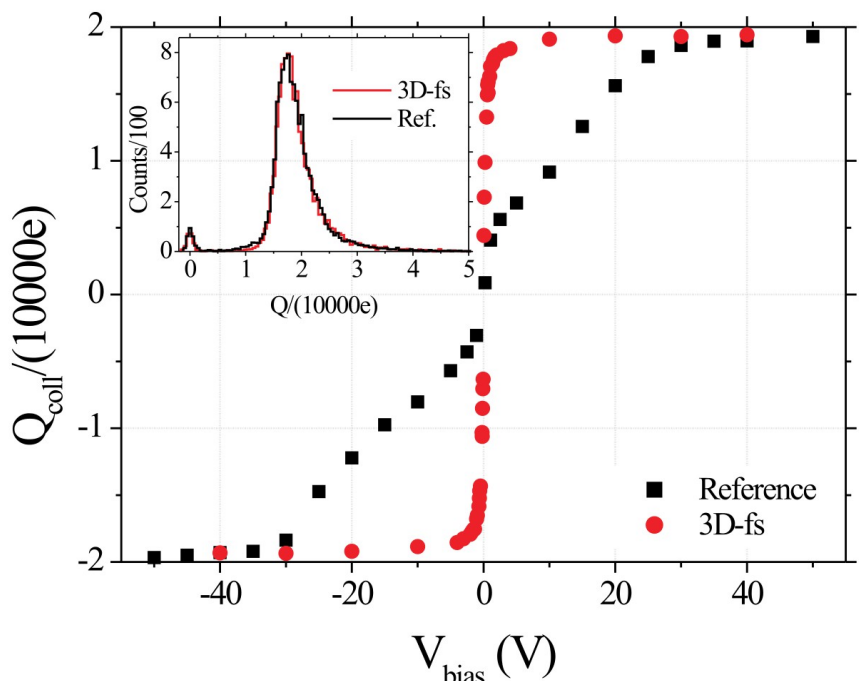

FIG. 5. Mean signal of two sensors fabricated on the same monocrystalline diamond, a reference conventional planar detector and a 3D-fs sensor in the IDC configuration. In the inset, the signal distribution from the two sensors at saturation voltage

An emerging feature, in all the sensors fabricated to date, is the lower response of the 3D-devices fabricated with the nanosecond laser, compared with the reference or with the corresponding fs-made structures fabricated in the same kind of diamond, justified in terms of the nanocrystalline $sp^3$ defective phase evidenced by Raman characterization [21] . The fs-columns are undoubtedly more efficient, but their electrical resistance is higher resulting in a higher Johnson noise in implemented 3D detector devices, which can be a substantial drawback. A better tuning of the graphitization parameters is required to minimize this defective layer.

**Conclusion**

All the expertise acquired in the fabrication and characterization of micro-structures in diamond can be considered ready to use for the realization of diamond integrated devices. Particularly, work is in progress to integrate horizontal and vertical graphitized structures fabricated with different techniques.